\pdfoutput=1
\RequirePackage{ifpdf}
\ifpdf
\documentclass[pdftex]{sigma}
\else
\documentclass{sigma}
\fi

\usepackage{cite}

\newcommand{\pr}{\operatorname{pr}}

\def\k{\kappa}
  \def\({\left(} \def\){\right)}
\def\bc{\begin{center}} \def\l{\label}
\def\ec{\end{center}}

\begin{document}
\allowdisplaybreaks

\renewcommand{\thefootnote}{$\star$}

\renewcommand{\PaperNumber}{001}

\FirstPageHeading

\ShortArticleName{Multi-Component Integrable Systems and Invariant Curve Flows in Certain
Geometries}

\ArticleName{Multi-Component Integrable Systems\\ and Invariant Curve Flows in Certain
Geometries\footnote{This
paper is a~contribution to the Special Issue ``Symmetries of Dif\/ferential Equations: Frames,
Invariants and Applications''.
The full collection is available
at
\href{http://www.emis.de/journals/SIGMA/SDE2012.html}{http://www.emis.de/journals/SIGMA/SDE2012.html}}}

\Author{Changzheng QU~$^\dag$ and Junfeng SONG~$^\ddag$ and Ruoxia YAO~$^\S$}

\AuthorNameForHeading{C.Z.~Qu, J.F.~Song and R.X.~Yao}

\Address{$^\dag$~Center for Nonlinear Studies, Ningbo University,
Ningbo, 315211, P.R.~China}
\EmailD{\href{mailto:quchangzheng@nbu.edu.cn}{quchangzheng@nbu.edu.cn}}

\Address{$^\ddag$~College of Mathematics and Information Science, Shaanxi Normal University,\\
\hphantom{$^\ddag$}~Xi'an, 710062, P.R.~China}
\EmailD{\href{mailto:songjunfeng1979@yahoo.com.cn}{songjunfeng1979@yahoo.com.cn}}

\Address{$^\S$~School of Computer Science, Shaanxi Normal University, Xi'an, 710062, P.R.~China}
\EmailD{\href{mailto:rxyao@snnu.edu.cn}{rxyao@snnu.edu.cn}}

\ArticleDates{Received September 28, 2012, in f\/inal form December 27, 2012; Published online January 02, 2013}

\Abstract{In this paper, multi-component generalizations to the Camassa--Holm equation,
the modif\/ied Camassa--Holm equation with cubic nonlinearity are introduced.
Geometric
formulations to the dual version of the Schr\"{o}dinger equation, the complex Camassa--Holm
equation and the multi-component modif\/ied Camassa--Holm equation are provided.
It is shown
that these equations arise from non-streching invariant curve f\/lows respectively in the
three-dimensional Euclidean geometry, the two-dimensional M\"{o}bius sphere
and $n$-dimensional sphere ${\mathbb S}^n(1)$.
Integrability to these systems is also studied.}


\Classification{37K10; 51M05; 51B10}

\renewcommand{\thefootnote}{\arabic{footnote}}
\setcounter{footnote}{0}

\section{Introduction}

Integrable systems solved by the inverse scattering method usually arise from
shallow water wave, physics, optical communication and applied
sciences etc.
Integrable systems have many interesting properties,
such as Lax-pair, inf\/inite number of conservation laws and
Lie--B\"{a}cklund symmetries, multi-solitons, B\"{a}cklund
transformations and bi-Hamiltonian structure etc.~\cite{abl,olv2}, which
are helpful to explore other properties of integrable systems~\cite{abl,olv2,tao}.

It is of great interest to study geometric aspects of integrable
systems.
So far, very few integrable systems were found to have
geometric formulations.
The relationship between completely integrable systems and the
f\/inite-dimensional dif\/ferential geometry of curves has been
studied extensively.
It turns out that some integrable systems arise from invariant curve
f\/lows in certain geometries~\cite{anc1, anc2, anc3, anc4, asa, cal, cho1, cho2, cho3, cho4, con1,
con2, dol, gol, gloq, has, ive, kou, lan, li, bef1, bef2, bef3, bef4, bef5, bef6, bef7, bef8, bef9,
mis,
mus, nak, olv1, olv4, pin, san1, qia1, san2, ter, sw, son2, wan, wo}.
The pioneering work on this
topic was done by Hasimoto~\cite{has}.
He showed that the
integrable nonlinear Schr\"{o}dinger equation (NLS)
\begin{gather*}
i\phi_t+\phi_{ss}+|\phi|^2\phi=0
\end{gather*}
is equivalent to the system for the curvature $\k$ and $\tau$ of curves $\gamma$ in ${\mathbb R}^3$
\begin{gather}\label{e1.1}
\k_t=-2\tau \k_s-\k\tau_s,\qquad
\tau_t=\frac{\k_{sss}}{\k}-\frac{\k_s\k_{ss}}{\k^2}-2\tau\tau_s+\k\k_s
\end{gather}
via the so-called Hasimoto
transformation $\phi=\k\exp(i\int^s\tau(t,z){\rm d} z)$. Indeed, the
system~\eqref{e1.1} is equivalent to the vortex f\/ilament equation
\begin{gather}
\gamma_t=\gamma_s\times \gamma_{ss}=\k {\bf b}, \l{e1.2}
\end{gather}
where ${\bf b}$ is the binormal
vector of $\gamma$.
Mar\'{i}~Bef\/fa, Sanders and Wang
\cite{bef1,san1} noticed that Hasimoto transformation is a~gauge transformation
relating the Frenet frame ${\bf\{t,n,b\}}$ to the parallel frame
$\{{\bf t}^1,{\bf n}^1,{\bf b}^1\}$.
It is also a~Poisson map which takes Hamiltonian structure
of the NLS equation to that of the vertex f\/ilament
f\/low~\cite{lan}.
The Hasimoto transformation has been generalized in \cite{san1} to the
Riemannian manifold with constant curvature, which is used to obtain the corresponding
integrable equations associated with the invariant non-stretching
curve f\/lows.
The parallel frames and other kinds of frames are also
used to derive bi-Hamiltonian operators and associated
hierarchies of multi-component soliton
equations from non-stretching curve f\/lows on Lie group
manifolds~\cite{anc1,anc3, bef1,bef9}.
The KdV equation, the
modif\/ied KdV equation, the Sawada--Kotera equation and the
Kaup--Kuperschmidt equation were shown to arise from the invariant
curve f\/lows respectively in centro-equiaf\/f\/ine geometry
\cite{cal,cho1,pin}, Euclidean geometry~\cite{gol}, special af\/f\/ine
geometry~\cite{cho4,bef5} and projective geometries~\cite{cho4, li, mus}.

The integrable systems with non-smooth solitary waves have drawn much
attention in the last two decades because of their remarkable properties.
The celebrated Camassa--Holm (CH) equation
\begin{gather}\label{e1.2.1}
m_t+um_x+2u_xm+a u_x=0,\qquad m=u-u_{xx},
\end{gather}
was proposed as a~model for the unidirectional propagation of the
shallow water waves over a~f\/lat bottom, with $u(x,t)$ representing
the water's free surface in non-dimensional variables~\cite{cam}.
It
was also found using the method of recursion operators by Fokas and
Fuchssteiner~\cite{fuc1} as a~bi-Hamiltonian equation with an
inf\/inite number of conserved functionals.
Geometrically, the Camassa--Holm equation
arises from a~non-stretching invariant planar curve f\/low in the
centro-equiaf\/f\/ine geometry~\cite{cho1}, and the periodic CH
equation~\eqref{e1.2.1} describes geodesic f\/lows on ${\rm dif\/f}({\mathbb{
S}}^1\times{\mathbb{R}})$
with respect to right-invariant Sobolev $H^1$ metric for $a=0$~\cite{con1,con2,kou}
and Bott--Virasoro algebra for $a\not=0$~\cite{mis}.
A dual version of the
Ito system is the two-component Camassa--Holm equation~\cite{olv3}, the periodic
two-component CH equation also describe geodesic f\/lows on an extended Bott--Virasoro
algebra~\cite{guh}.

It is remarked that all nonlinear terms in the CH equation
are quadratic.
In contrast to the integrable modif\/ied KdV equation
with a~cubic nonlinearity, it is of great interest to f\/ind
integrable CH-type equations with cubic or higher-order nonlinearity
and non-smooth solitary waves.
To the best of our knowledge,
two scalar integrable CH-type equations with cubic nonlinearity have been
discovered.
The f\/irst equation is~\cite{fuc2,olv3,qia1}
\begin{gather}\label{e1.3}
m_t+\frac12\left(\big(u^2+\delta u_x^2\big)m\right)_x=0,\qquad m=u+\delta u_{xx},
\end{gather}
where $\delta=\pm1$, and the second one is the so-called Novikov
equation~\cite{hon, nov}
\begin{gather*}
m_t+3uu_xm+u^2m_x=0,\qquad m=u-u_{xx},
\end{gather*}
which are completely integrable, and admit peaked solitons.
Recently, systems of
CH-type equations with cubic nonlinearity were also obained~\cite{gen, son2}.

The CH equation can also be derived by the
tri-Hamiltonian duality approach basing on bi-Hamiltonian structure of the KdV equation.
Other examples of dual integrable systems obtained using the method of tri-Hamiltonian duality can
be found in
\cite{fuc2,olv3}.
Nonlinear dual integrable systems, such as the CH equation and the modif\/ied CH
equations, are endowed with nonlinear dispersion, which in most
cases, enables these systems to support non-smooth soliton-like structures.
It was remarked in~\cite{gloq} that the modif\/ied
CH equation~\eqref{e1.3} can be regarded as a~Euclidean-invariant
version of the CH equation~\eqref{e1.2.1}, just as the modif\/ied KdV equation is a~Euclidean-invariant counterpart to the KdV equation from the
viewpoint of curve f\/lows in Klein geometries~\cite{cho1,cho2,gol,pin}.

The aim of this paper is to provide geometric formulations to multi-component integrable systems
admitting non-smooth solitons.
We shall show that several multi-component integrable
systems with non-smooth solitons, such as a~dual version of the Schr\"{o}dinger equation
\cite{fok,olv3}, the complex CH equation and multi-component modif\/ied CH equations arise
from the invariant curve f\/lows respectively in three-dimensional Euclidean geometry,
M\"{o}bius sphere and the $n$-dimensional unit sphere~${\mathbb{S}}^n(1)$.
To obtain integrable
systems relating to these geometric f\/lows, we shall use the scale limit technique.
The outline of this paper is as follows.
In Section~\ref{section2}, a~non-stretching invariant binormal
curve f\/low in ${\mathbb R}^3$ is introduced and studied.
Making use of the system for
curvature and torsion corresponding to this f\/low, we obtain a~novel integrable Schr\"{o}dinger
equation by a~scale limit approach, which is completely integrable system and can be obtained
by the so-called tri-Hamiltonian duality approach~\cite{fuc2,olv3}.
In Section~\ref{section3}, we give a~brief discussion on M\"{o}bius 2-sphere ${\rm PO}(3,1)/H$ and the $n$-dimensional
sphere ${\rm SO}(n+1)/{\rm SO}(n)$,
the Cartan structure equations for curves in both geometries are reviewed, which will
be used in subsequent sections to study curve f\/lows in both geometries.
In Section~\ref{section4}, we
consider the non-stretching curve f\/lows in M\"{o}bius 2-sphere.
It is shown that the complex
Camassa--Holm equation and complex Hunter--Saxton equation describe the non-stretching curve f\/lows
in M\"{o}bius 2-sphere.
The bi-Hamiltonian structure for the complex Camassa--Holm equation is obtained.
In Section~\ref{section5}, we study non-stretching curve f\/lows in the $n$-dimensional sphere ${\mathbb
S}^n(1)$.
Interestingly, we f\/ind that a~multi-component modif\/ied CH equation (a multi-component
generalization
of the modif\/ied Camassa--Holm equation) is equivalent to a~non-stretching curve f\/low in ${\mathbb
S}^n(1)$.
Integrability of the system is identif\/ied.
Finally, Section~\ref{section6} contains concluding remarks on this work.

\section{An integrable nonlinear Schr\"{o}dinger equation}\label{section2}

We consider the f\/lows of space curves in ${\mathbb R}^3$, given by
\begin{gather}\label{S-Flow}
\gamma_t=U{\bf{n}}+V{\bf{b}}+W{\bf{t}},
\end{gather}
where $\bf{t}$, $\bf{n}$ and $\bf{b}$ denote the tangent, normal and binormal vectors of the curves,
respectively.
The velocities $U$, $V$ and $W$ depend on the curvature and torsion as well as their
derivatives with respect to arc-length parameter $s$.
The arc-length parameter $s$ is def\/ined
implicitly by ${\rm d} s=h{\rm d} p$, $h=|\gamma'(p)|$, where $p$ is a~free parameter and is independent of
time.
We denote by $\kappa$ and $\tau$ the curvature and torsion of the curves, respectively.
Governed by the f\/low~\eqref{S-Flow}, time evolutions of those geometric invariants are given
by~\cite{has,nak}
\begin{gather}
\dot{\bf{t}}=\left(\frac{\partial U}{\partial s}-\tau V
+\kappa W\right){\bf{n}}+\left(\frac{\partial V}{\partial s}+\tau U\right){\bf{b}},\nonumber\\
\dot{\bf{n}}=-\left(\frac{\partial U}{\partial s}-\tau V+\kappa W\right){\bf{t}}+
\left[\frac{1}{\kappa}\frac{\partial}{\partial s}\left(\frac{\partial V}{\partial s}
+\tau U\right)+\frac{\tau}{\kappa}\left(\frac{\partial U}{\partial s}
-\tau V+\kappa W\right)\right]{\bf{b}},\nonumber\\
\dot{\bf{b}}=-\left(\frac{\partial V}{\partial s}+\tau U\right){\bf{t}}-
\left[\frac{1}{\kappa}\frac{\partial}{\partial s}\left(\frac{\partial V}{\partial s}
+\tau U\right)+\frac{\tau}{\kappa}\left(\frac{\partial U}{\partial s}
-\tau V+\kappa W\right)\right]{\bf{n}},\nonumber\\
\dot{h}=2h\left(\frac{\partial W}{\partial s}-\kappa U\right)\label{G-Evolution}
\end{gather}
and
\begin{gather}
\frac{\partial\tau}{\partial t}=\frac{\partial}{\partial s}\left[\frac{1}
{\kappa}\frac{\partial}{\partial s}\left(\frac{\partial V}{\partial s}
+\tau U\right)+\frac{\tau}{\kappa}\left(\frac{\partial U}{\partial s}-
\tau V\right)+\tau\int^s\kappa U{\rm d} s'\right]+\kappa\tau U+\kappa\frac{\partial V}{\partial
s},\nonumber\\
\frac{\partial\kappa}{\partial t}=\frac{\partial^2U}{\partial s^2}
+\big(\kappa^2-\tau^2\big)U+\frac{\partial\kappa}{\partial s}\int^s\kappa U{\rm d} s'
-2\tau\frac{\partial V}{\partial s}-\frac{\partial\tau}{\partial s}V.\label{KT-Evolution}
\end{gather}

Assuming that the f\/low is intrinsic, namely the arc-length does not depend on time, it implies from
\eqref{G-Evolution} that
\begin{gather}\label{W-U}
W_s=\kappa U.
\end{gather}
In terms of~\eqref{KT-Evolution}, one f\/inds that the complex function
\begin{gather*}
\phi=\kappa\eta,\qquad\eta=\exp\left[i\int^s\tau(s',t){\rm d} s'\right]
\end{gather*}
satisf\/ies the equation~\cite{has,nak}
\begin{gather}
\phi_t=\left(\frac{\partial^2}{\partial s^2}+|\phi|^2+i\phi\int^s {\rm d} s'
\tau\bar{\phi}+\frac{\partial\phi}{\partial s}\int^s {\rm d} s'\bar{\phi}\right)(U\eta) \nonumber\\
\hphantom{\phi_t=}{} +\left(i\frac{\partial^2}{\partial s^2}+i|\phi|^2+\phi\int^s {\rm d} s'
\tau\bar{\phi}-i\phi\int^s {\rm d} s'\frac{\partial\bar{\phi}}{\partial s'}\right)(V\eta),\label{Hasimoto-Evolution}
\end{gather}
where $\bar{\phi}$ denotes the complex conjugate of $\phi$.

Let $U=0$, $V=\kappa$, where $\kappa$ is a~real function, then~\eqref{W-U}
implies that $W=C_1$, where $C_1$ is a~constant.
Setting $C_1=0$, we derive
from~\eqref{Hasimoto-Evolution} the celebrated Schr\"{o}dinger equation
\begin{gather}\label{Schrodinger}
i\phi_t+\phi_{ss}+\frac12|\phi|^2\phi=0.
\end{gather}

Let $U=-\kappa_s$, $V=-\kappa\tau$, then $W=-\frac12\kappa^2+C_2$,
where $C_2$ is a~constant.
Letting $C_2=0$, we f\/ind that $\phi$ satisf\/ies the mKdV system
\begin{gather*}
\phi_t+\phi_{sss}+\frac32|\phi|^2\phi_s=0.
\end{gather*}

In the following, we shall consider the case $U=W=0$.
Denote $\theta(s,t)=\int^s\tau(s',t){\rm d} s'$,
$g=V\eta$.
It follows from~\eqref{Hasimoto-Evolution} that $\phi$ satisf\/ies the equation
\begin{gather}\label{Cur-System}
i\phi_t+g_{ss}+|\phi|^2g-\phi\int^sg(\cos\theta-i\sin\theta)k_{s'}{\rm d} s'=0.
\end{gather}

Set $\tilde{u}=\kappa\cos\theta$, $\tilde{v}=\kappa\sin\theta$, $g=g_1+ig_2$,
then the equation~\eqref{Cur-System} is separated to two equations
\begin{gather*}
\tilde{u}_t=-g_{2,ss}-\tilde{v}\partial_s^{-1}[\kappa(g_1\cos\theta+g_2\sin\theta)_s],\nonumber\\
\tilde{v}_t=g_{1,ss}+\tilde{u}\partial_s^{-1}[\kappa(g_1\cos\theta+g_2\sin\theta)_s].
\end{gather*}

Furthermore, letting $\tilde{u}=u+v_s$, $\tilde{v}=v-u_s$, and choosing the
binormal velocity $V$ to be $V=\partial_s^{-1}[\big(u^2+v^2\big)_s/\kappa]$,
we f\/ind that $u$ and $v$ satisfy the system
\begin{gather}
(u+v_s)_t=-g_{2,ss}-(v-u_s)\big(u^2+v^2\big),\nonumber\\
(v-u_s)_t=g_{1,ss}+\big(u^2+v^2\big)(u+v_s)\label{Sch-System}
\end{gather}
with
\begin{gather*}
g_1=\frac{(u+v_s)\big(u^2+v^2\big)_s}{(u+v_s)^2+(v-u_s)^2},\qquad g_2
=\frac{(v-u_s)\big(u^2+v^2\big)_s}{(u+v_s)^2+(v-u_s)^2}.
\end{gather*}

Applying the scaling transformations
\begin{gather*}
s\longmapsto s,\qquad
t\longmapsto\epsilon^2t,\qquad
u\longmapsto\epsilon^{-1}u,\qquad
v\longmapsto\epsilon^{-1}v
\end{gather*}
to system~\eqref{Sch-System} produces
\begin{gather}
(u+v_s)_t=-\epsilon^2g_{2,ss}-(v-u_s)\big(u^2+v^2\big),\nonumber\\
(v-u_s)_t=\epsilon^2g_{1,ss}+\big(u^2+v^2\big)(u+v_s).\label{Sch-System-1}
\end{gather}
Expanding $u$ and $v$ in powers of the small parameter $\epsilon$
\begin{gather*}
u(t,s)=u_0(t,s)+\epsilon u_1(t,s)+\epsilon^2u_2(t,s)+\>\cdots,\\
v(t,s)=v_0(t,s)+\epsilon v_1(t,s)+\epsilon^2v_2(t,s)+\>\cdots,
\end{gather*}
and plugging them into system~\eqref{Sch-System-1}, we f\/ind
that the leading order terms $u_0(t,s)$ and $v_0(t,s)$ satisfy the system
\begin{gather}
(u_0+v_{0,s})_t+\big(u_0^2+v_0^2\big)(v_0-u_{0,s})=0,\nonumber\\
(v_0-u_{0,s})_t-\big(u_0^2+v_0^2\big)(u_0+v_{0,s})=0.\label{Sch-System-2}
\end{gather}
Again we use the notation $\phi(t,s)=u_0(t,s)+iv_0(t,s)$,
then it is inferred from~\eqref{Sch-System-2} that $\phi(t,s)$ satisf\/ies the equation
\begin{gather}\label{Dual-Schrodinger}
i(\phi_t-i\phi_{ts})+|\phi|^2(\phi-i\phi_s)=0,
\end{gather}
which is a~dual version of the Schr\"{o}dinger equation~\eqref{Schrodinger},
and can be obtained by the approach of tri-Hamiltonian duality~\cite{fuc2,olv3}.
Equation~\eqref{Dual-Schrodinger} is formally completely integrable since it
admits bi-Hamiltonian structure~\cite{olv3}
\begin{gather*}
\rho_t=\mathcal{E}_1\frac{\delta H_2}{\delta\rho}=\mathcal
{E}_2\frac{\delta H_1}{\delta\rho},
\end{gather*}
where $\rho=\phi-i\phi_s$, $\mathcal{E}_1$ and $\mathcal{E}_2$
def\/ined by
\begin{gather*}
\mathcal{E}_1=\partial_s+i\qquad\text{and}\qquad\mathcal{E}_2(F)=\rho
\partial_s^{-1}(\bar{\rho}F-\rho\bar{F})
\end{gather*}
are compatible Hamiltonian operators, while
\begin{gather*}
H_1=\int_{\mathbb{R}}\left(|\phi|^2-i\bar{\phi}\phi_s\right){\rm d} s=\int_{\mathbb{R}}\rho\bar{\phi}{\rm d} s,
\qquad
H_2=\frac1{2}\int_{\mathbb{R}}|\phi|^2\left(|\phi|^2-i\bar{\phi}\phi_s\right){\rm d} s.
\end{gather*}
are the corresponding Hamiltonian functionals.

\section[M\"{o}bius sphere ${\rm PO}(3,1)/H$ and unit sphere ${\rm SO}(n+1)/{\rm SO}(n)$]{M\"{o}bius sphere $\boldsymbol{{\rm PO}(3,1)/H}$ and unit sphere $\boldsymbol{{\rm SO}(n+1)/{\rm SO}(n)}$}\label{section3}

In this section, we give a~brief account of M\"{o}bius 2-sphere
${\rm PO}(3,1)/H$ and unit sphere $S^n(1)={\rm SO}(n+1)/{\rm SO}(n)$.
Please refer to
the book~\cite{sha} for the details of the two geometries.

\subsection{M\"{o}bius 2-sphere}

Let $(u_0,u_1,u_2,u_3)\in{\mathbb R}^4$, we def\/ine the inner product on ${\mathbb R}^4$ by
\begin{gather*}
\langle x,y\rangle =x^T\Lambda_{3,1}y,
\end{gather*}
where $x,y\in{\mathbb R}^4$, and the matrix $\Lambda_{3,1}$ is
\begin{gather*}
\Lambda_{3,1}=\left(
\begin{matrix}
0&0&0&-1\\
0&1&0&0\\
0&0&1&0\\
-1&0&0&0
\end{matrix}\right).
\end{gather*}
A vector f\/ield $x\in{\mathbb R}^4$ is said to be light-like, if it satisf\/ies $\langle x,x\rangle =0$.
All
the light-like vector f\/ields form a~set $L$, which is called optical cone, def\/ined by the equation
\begin{gather*}
2x_0x_3-x_1^2-x_2^2=0.
\end{gather*}
Clearly it is homogeneous, namely for any $\lambda\in{\mathbb R}$,
if $x\in L$, then $\lambda x\in L$.
The projectivisation of $L$ is said to be M\"{o}bius 2-sphere, which is isomorphic
to ${\mathbb S}^2$.
Recall that
\begin{gather*}
{\rm O}(3,1)=\left\{g\in {\rm GL}(4,{\mathbb R}):g^T\Lambda_{3,1}g=\Lambda_{3,1}\right\},
\end{gather*}
and the M\"{o}bius group is def\/ined to be ${\rm PO}(3,1)={\rm O}(3,1)/{\pm I}$.
We denote
\begin{gather*}
H=\left\{h\in {\rm PO}(3,1):h[e_3]=[e_3]\right\}\\
\phantom{H}
=\left\{\left(
\begin{matrix}
a^{-1}&0&0\\
v&A&0\\
b&\eta^T & a
\end{matrix}
\right)
\in {\rm O}(3,1); \; A\in {\rm O}(2), \; a\in{\mathbb R}^+, \; v\in{\mathbb R}^2\right\},
\end{gather*}
where $[e_3]$ denotes the equivalent class of $e_3$ in $P({\mathbb
R}^4)$, $[e_3]=(0,0,0,\ast)$.
A straightforward
computation gives
\begin{gather*}
\eta=aA^Tv,\qquad b=\frac a2v^Tv=\frac a2\|v\|^2.
\end{gather*}
It is easy to verify that the group ${\rm PO}(3,1)$ acts on the M\"{o}bius
sphere transitively (the group action is the usual conformal
transformation).

\begin{definition}
The Klein pair $({\rm PO}(3,1),H)$ is called the M\"{o}bius 2-sphere.
\end{definition}

For any $g\in {\rm PO}(3,1)$, there exists a~unique decomposition around the identity of the group
\begin{gather*}
g=g_1g_0g_{-1}=\left(
\begin{matrix}
1&0&0\\
v&I_2&0\\
\frac12\|v\|^2&v^T & 1
\end{matrix}
\right)\left(
\begin{matrix}
a^{-1}&0&0\\
0&A&0\\
0&0&a
\end{matrix}
\right)\left(
\begin{matrix}
1&u^T&\frac12\|u\|^2\\
0&I_2&u\\
0&0&1
\end{matrix}
\right),
\end{gather*}
where $a\in{\mathbb R}^+$, $A\in {\rm O}(2)$, $u,v\in{\mathbb R}^2$,
$h=g_1g_0\in H$, $u$ in $g_{-1}$ part represent a~local coordinate
of the point in ${\rm PO}(3,1)/H$.
For the corresponding Lie algebra
${\mathfrak g}$, there exists also a~decomposition
\begin{gather*}
{\mathfrak o}(3,1)={\mathfrak g}={\mathfrak g}_1\oplus{\mathfrak
g}_0\oplus{\mathfrak g}_{-1},
\end{gather*}
where
\begin{gather*}
\left(
\begin{matrix}
0&0&0\\
p&0&0\\
0&p^T&0
\end{matrix}
\right)
\in{\mathfrak g}_1,
\qquad\left(
\begin{matrix}
-\epsilon&0&0\\
0&S&0\\
0&0&\epsilon
\end{matrix}
\right)
\in{\mathfrak g}_0,
\qquad\left(
\begin{matrix}
0&q^T&0\\
0&0&q\\
0&0&0
\end{matrix}
\right)
\in{\mathfrak g}_{-1},
\end{gather*}
$p,q\in{\mathbb R}^2$, $\epsilon\in{\mathbb R}$, $S\in {\rm O}(2)$.
The
Lie algebra of the isotropy group $H$ is ${\mathfrak
h}={\mathfrak g}_1\oplus{\mathfrak g}_0$ while ${\mathfrak
g}/{\mathfrak h}={\mathfrak g}_{-1}$ is identif\/ied to the tangent
space of the conformal sphere ${\rm PO}(3,1)/H$.

\subsection[$n$-dimensional sphere ${\mathbb S}^n(1)={\rm SO}(n+1)/{\rm SO}(n)$]{$\boldsymbol{n}$-dimensional sphere $\boldsymbol{{\mathbb S}^n(1)={\rm SO}(n+1)/{\rm SO}(n)}$}

The $n$-dimensional unit-sphere is also a~homogeneous space $M=G/H={\rm SO}(n+1)/{\rm SO}(n)$.
The corresponding Lie algebra has the following Cartan--Killing decomposition
\begin{gather*}
\mathfrak{so}(n+1)={\mathfrak h}\oplus{\mathfrak
m}=\mathfrak{so}(n)\oplus{\mathbb R}^n,
\end{gather*}
with
\begin{gather*}
\left(
\begin{matrix}
0&-p^T\\
p&0
\end{matrix}
\right)\in{\mathfrak m},\qquad\left(
\begin{matrix}
0&0\\
0&\Theta
\end{matrix}
\right)\in{\mathfrak h},
\end{gather*}
where $p\in{\mathbb R}^n$, $\Theta\in\mathfrak{so}(n)$, and the
decomposition satisf\/ies
\begin{gather*}
[{\mathfrak h},{\mathfrak h}]\subset{\mathfrak h},\qquad
[{\mathfrak h},{\mathfrak m}]\subset{\mathfrak m},\qquad
[{\mathfrak m},{\mathfrak m}]\subset{\mathfrak h},
\end{gather*}
where ${\mathfrak m}$ is identif\/ied to the tangent space $T_xM\cong
{\mathbb R}^n$ of $M={\rm SO}(n+1)/{\rm SO}(n)$.
The f\/lat Cartan connection of
principle ${\rm SO}(n)$ bundle ${\rm SO}(n+1)\rightarrow{\mathbb S}^n$ is
given by the Maurer--Cartan form of Lie group ${\rm SO}(n+1)$.
The Cartan
structure equation reads as
\begin{gather*}
\Omega={\rm d}\omega+\frac12[\omega,\omega]=0,
\end{gather*}
where $\mathfrak{so}(n+1)$-valued one-form $\omega$ is
decomposed to
\begin{gather*}
\omega=\omega_H+\theta,\qquad\omega_H\in\Lambda^1(P,{\mathfrak
h}),\qquad\theta\in\Lambda^1(P,{\mathfrak g}/{\mathfrak h}),
\end{gather*}
where ${\mathfrak g}/{\mathfrak h}$-valued $\theta$ represents a
linear coframe, ${\mathfrak h}$-valued $\omega_H$ represents a
linear connection on~${\mathbb S}^n$.
The corresponding Cartan
structure equation is separated to
\begin{gather*}
{\mathfrak J}\equiv {\rm d}\theta+\frac12[\omega_H,\theta]+\frac12[\theta,\omega_H]=0
\end{gather*}
and
\begin{gather*}
{\mathfrak R}\equiv {\rm d}\omega_H+\frac12[\omega_H,\omega_H]=-\frac12[\theta,\theta],
\end{gather*}
where ${\mathfrak J}$ and ${\mathfrak R}$ are called torsion and curvature forms, respectively.

\section[Curve flows in ${\rm PO}(3,1)/H$ and the complex CH equation]{Curve f\/lows in $\boldsymbol{{\rm PO}(3,1)/H}$ and the complex CH equation}\label{section4}

For the M\"{o}bius geometry ${\rm PO}(3,1)/H$, its Cartan connection takes
values on ${\mathfrak g}=o(3,1)$, with the form
\begin{gather*}
\omega=\left(
\begin{matrix}
-\epsilon&\xi^T&0\\
\eta&\Theta&\xi\\
0&\eta^T&\epsilon
\end{matrix}
\right),
\end{gather*}
and the Cartan structure equation reads as
\begin{gather*}
\Omega={\rm d}\omega+\frac12[\omega,\omega]=0.
\end{gather*}

Let's consider the invariant curve f\/lows for curves
$\gamma(x,t)=(u_1(x,t),u_2(x,t))$ on the conformal sphere
$M^2={\rm PO}(3,1)/H$, where $x$ denotes the parameter of the curves, $t$
is the time variable, $u_1$ and $u_2$ denote the local coordinates on
$M^2$.
Let $\gamma_t=\gamma_{\ast}\frac{\partial}{\partial_t}$ be the evolutionary vector f\/ield of the
curves, $\gamma_x=\gamma_{\ast}\frac{\partial}{\partial_x}$ denotes the tangent vector of the
curves.
Assuming
that the curve f\/low is intrinsic, namely the parameter $x$ for the
curve does not depend on time $t$, we have
\begin{gather*}
[\gamma_x,\gamma_t]=0.
\end{gather*}
It was shown in~\cite{fel,bef2} that there exists a~conformally
equivariant moving frame (the Frenet frame) $\rho=\rho(x,t)\in {\rm PO}(3,1)$
along the curve $\gamma(x,t)\subset M^2$.
Let $D_x$ and $D_t$ denote
respectively the vector f\/ield $\frac{d}{dx}$ and $\frac{d}{dt}$ along
the curves $\rho$ in ${\rm PO}(3,1)$, then the Frenet formulae for the conformally parametric curves is
\begin{gather*}
\rho_x=\rho\hat{\omega}(D_x),
\end{gather*}
with
\begin{gather}\label{Frenet-Matrix}
\hat{\omega}(D_x)=\left(
\begin{matrix}
0&1&0&0\\
k_1&0&0&1\\
k_2&0&0&0\\
0&k_1&k_2&0
\end{matrix}
\right),
\end{gather}
where $k_1$ and $k_2$ are the conformally dif\/ferential invariants for the curve $\gamma(x,t)$.
The time evolution for the frame $\rho(x,t)$ can be written as
\begin{gather*}
\rho_t=\rho\hat{\omega}(D_t),
\end{gather*}
where
\begin{gather}\label{Evolution-Matrix}
\hat{\omega}(D_t)=\left(
\begin{matrix}
-\epsilon&h_1&h_2&0\\
f_1&0&-\alpha&h_1\\
f_2&\alpha&0&h_2\\
0&f_1&f_2&\epsilon
\end{matrix}\right),
\end{gather}
and $\epsilon$, $\alpha$, $f_i$, $h_i$ $(i=1,2)$ are some conformal dif\/ferential
invariants related to $k_1$ and $k_2$, to be determined.
By the Cartan structure equation, one gets
\begin{gather}\label{Cratan-SEquation}
\Omega(D_x,D_t)=\frac{d}{dt}\hat{\omega}(D_x)-\frac{d}{dx}\hat{\omega}(D_t)-[\hat{\omega}(D_x),\hat{\omega}(D_t)]=0.
\end{gather}
Plugging~\eqref{Frenet-Matrix} and \eqref{Evolution-Matrix} into
\eqref{Cratan-SEquation} results in the following equations
\begin{gather}
\epsilon=-h_{1,x},\qquad\alpha=h_{2,x},\label{Constraint-1}\\
f_1=\epsilon_x+k_1h_1+k_2h_2,\label{Constraint-2}\\
f_2=\alpha_x+k_2h_1-k_1h_2,\label{Constraint-3}\\
k_{1,t}=f_{1,x}-k_1\epsilon+\alpha k_2,\label{Constraint-4}\\
k_{2,t}=f_{2,x}-\epsilon k_2-\alpha k_1,\label{Constraint-5}
\end{gather}
where~\eqref{Constraint-1} is the torsion part (i.e.,the ${\mathfrak
g}_{-1}$ part) of the Cartan structure equation
\eqref{Cratan-SEquation}, which can be written as
\begin{gather*}
\left(
\begin{matrix}\epsilon\\
\alpha
\end{matrix}\right)=\left(
\begin{matrix}
-\partial_x\\
\partial_x
\end{matrix}
\right)\left(
\begin{matrix}h_1\\
h_2
\end{matrix}\right)\equiv J_1\left(
\begin{matrix}h_1\\
h_2
\end{matrix}\right).
\end{gather*}
Inserting~\eqref{Constraint-1} into \eqref{Constraint-2} and
\eqref{Constraint-3} gives
\begin{gather}\label{Constraint-6}
f_1=-h_{1,xx}+k_1h_1+k_2h_2,\qquad
f_2=h_{2,xx}+k_2h_1-k_1h_2.
\end{gather}
Substituting~\eqref{Constraint-1} and \eqref{Constraint-6} into \eqref{Constraint-4} and
\eqref{Constraint-5}, we obtain the evolution equations for the curvatures $k_1$
and $k_2$~\cite{bef2,bef4}
\begin{gather*}
k_{1,t}=-h_{1,xxx}+2k_1h_{1,x}+k_{1,x}h_1+2k_2h_{2,x}+k_{2,x}h_2,\\
k_{2,t}=h_{2,xxx}-2k_1h_{2,x}-k_{1,x}h_2+2k_2h_{1,x}+k_{2,x}h_1,
\end{gather*}
which is equivalent to
\begin{gather}\label{Curvature-Equation}
\left(
\begin{matrix}
k_{1}\\
k_2\end{matrix}\right)_t=\left(
\begin{matrix}
-\partial^3+k_1\partial+\partial k_1 &k_2\partial+\partial k_2\\
k_2\partial+\partial k_2&\partial^3-k_1\partial-\partial k_1
\end{matrix}\right)
\left(
\begin{matrix}h_1\\ h_2
\end{matrix}\right)\equiv J_2
\left(
\begin{matrix}h_1\\ h_2
\end{matrix}\right).
\end{gather}
The following cases are considered.

  {\bf Case 1.} Setting $k_1=1/2+m\equiv1/2+u-u_{xx}$, $k_2=n\equiv v-v_{xx}$,
$h_1=1-u$ and $h_2=v$ in~\eqref{Curvature-Equation}, we obtain the new two-component CH equation
\begin{gather}
m_t+2u_x m+u m_x-2v_x n-v n_x=0,\nonumber\\
n_t+2u_x n+u n_x+2v_x m+v m_x=0.\label{Two-CH}
\end{gather}
The above system admits the following Lax-pair
\begin{gather}\label{Lax-Pair}
\phi_x=U\phi,\qquad\phi_t=V\phi,
\end{gather}
with
\begin{gather*}
U=\left(
\begin{matrix}
0&\frac 12 +\lambda m&\lambda n&0\\
1&0&0&\frac 12 +\lambda m\\
0&0&0&\lambda n\\
0&1&0&0
\end{matrix}
\right)
\qquad \text{and}\qquad
V=\left(
\begin{matrix}
-u_x&f_1&f_2&0\\
\lambda^{-1}-u&0&v_x&f_1\\
v&-v_x&0&f_2\\
0&\lambda^{-1}-u&v&u_x
\end{matrix}\right),
\end{gather*}
where $f_1=\frac12(\lambda^{-1}+u)+\lambda(vn-um)$, $f_2=\frac12v-\lambda(un+vm)$.

 {\bf Case 2.} Setting $k_1=m=-u_{xx}$, $k_2=n=-v_{xx}$, $h_1=1-u$,
$h_2=v$, we arrive at the complex Hunter--Saxton equation
\begin{gather*}
m_t+2u_x m+u m_x-2v_x n-v n_x=0,\nonumber\\
n_t+2u_x n+u n_x+2v_x m+v m_x=0,
\end{gather*}
which admits the Lax-pair~\eqref{Lax-Pair}
with
\begin{gather*}
U=\left(
\begin{matrix}
0&-\lambda u_{xx}&-\lambda v_{xx}&0\\
1&0&0&-\lambda u_{xx}\\
0&0&0&-\lambda v_{xx}\\
0&1&0&0
\end{matrix}
\right)
\end{gather*}
and
\begin{gather*}
V=\left(
\begin{matrix}
-u_x&\lambda(vn-um)&-\lambda(un+vm)&0\\
\lambda^{-1}-u&0&v_x&\lambda(vn-um)\\
v&-v_x&0&-\lambda (un+vm)\\
0&\lambda^{-1}-u&v&u_x
\end{matrix}
\right).
\end{gather*}

  {\bf Case 3.} In~\eqref{Curvature-Equation}, letting $h_1$ and $h_2$ satisfy
\begin{gather*}
k_1=\frac12\frac{h_1^2-h_2^2}{\big(h_1^2+h_2^2\big)^2},\qquad k_2=-\frac{h_1h_2}{\big(h_1^2+h_2^2\big)^2},
\end{gather*}
we get the two-component Harry--Dym equation
\begin{gather*}
h_{1,t}=\big(h_1^3-3h_1h_2^2\big)h_{1,xxx}+\big(h_2^3-3h_1^2h_2\big)h_{2,xxx},\\
h_{2,t}=\big(h_2^3-3h_1^2h_2\big)h_{1,xxx}-\big(h_1^3-3h_1h_2^2\big)h_{2,xxx}.
\end{gather*}

\begin{remark}
{\rm Mar\'i~Bef\/fa~\cite{bef2,bef4} showed that the complex KdV equation
arises from the invariant curve motion in M\"{o}bius 2-sphere.
Indeed, taking $h_1=-k_1$, $h_2=-k_2$ in~\eqref{Curvature-Equation} yields the complex KdV equation
\begin{gather*}
k_{1,t}=k_{1,xxx}-3k_1k_{1,x}+3k_2k_{2,x},\qquad
k_{2,t}=k_{2,xxx}-3k_1k_{2,x}-3k_2k_{1,x}.
\end{gather*}
Its Hamiltonian structure $J_2$, 
see~\eqref{Curvature-Equation}, 
was originally derived in~\cite{bef4}.
One can see that the
bi-Hamiltonian structure $J_1$ and $J_2$ of the complex KdV equation
comes from the Cartan structure equation for the conformal invariant
curve f\/low.
According to the decomposition of the Lie algebra
\begin{gather*}
{\mathfrak g}={\mathfrak g}_1\oplus{\mathfrak g}_0\oplus{\mathfrak
g}_{-1},
\end{gather*}
the Cartan curvature form $\Omega$ is decomposed to
$\Omega=\Omega_1+\Omega_0+\Omega_{-1}$, where $J_1$ comes from
torsion part of the structure equation, i.e., the $\Omega_{-1}$
part, and $J_2$ arises from the $\Omega_1$ part.
In the sequel, we will show that the complex CH equation admits a~bi-Hamiltonian structure
$\hat{\mathcal{J}}_1$ and $\hat{\mathcal{J}}_2$.
It turns out that the complex CH
equation is a~dual version of the complex KdV equation (in the sense of~\cite{olv3}).}
\end{remark}

It is well-known that the CH equation is a~bi-Hamiltonian system~\cite{cam}
\begin{gather*}
m_t=\mathcal{J}\frac{\delta H_1}{\delta m}=\mathcal{D}\frac{\delta H_2}{\delta m},
\end{gather*}
where $\mathcal{J}=-(m\partial+\partial m)$ and $\mathcal{D}=-(\partial-\partial^3)$
are Hamiltonian operators, $H_1=-\frac12\int um{\rm d} x$ and $H_2=-\frac12\int u\big(u^2+u_x^2\big){\rm d} x$
are the corresponding Hamiltonian functionals.
As for the CH equation, the complex CH
equation can be obtained by the approach of tri-Hamiltonian duality~\cite{olv3}.
Indeed, we have the following result.

\begin{theorem}
The complex CH equation~\eqref{Two-CH} is a~bi-Hamiltonian system, which can be written as
\begin{gather*}
\left(
\begin{matrix}m\\
n\end{matrix}\right)_t=\hat{\mathcal{J}}_1\left(
\begin{matrix}\dfrac{\delta\hat{H}_2}{\delta
m}\vspace{1mm}\\
\dfrac{\delta\hat{H}_2}{\delta n}\end{matrix}\right)=\hat{\mathcal{J}}_2\left(
\begin{matrix}
\dfrac{\delta\hat{H}_1}{\delta m}\vspace{1mm}\\ \dfrac{\delta\hat{H_1}}{\delta n}
\end{matrix}\right)
\end{gather*}
with
\begin{gather*}
\hat{\mathcal{J}}_1=\left(
\begin{array}{cc}
\partial^3-\partial&0\\
0&\partial-\partial^3
\end{array}\right),\qquad
\hat{\mathcal{J}}_2=\left(
\begin{matrix}
m\partial+\partial m&n\partial+\partial n\\
n\partial+\partial n&-m\partial -\partial m
\end{matrix}\right)
\end{gather*}
and
\begin{gather*}
\hat{H}_1=\frac12\int(vn-um){\rm d} x,\qquad\hat{H}_2=\frac12\int
\left[u\big(u^2+u_x^2\big)-u\big(3v^2+v_x^2\big)-2vu_xv_x\right]{\rm d} x.
\end{gather*}
\end{theorem}

\begin{proof} Clearly, $\hat{\mathcal{J}}_1$ and $\hat{\mathcal{J}}_2$ are skew symmetric.
To prove they are Hamiltonian
operators, it suf\/f\/ices to prove that the Poisson bracket def\/ined by $\hat{\mathcal{J}}_2$ satisf\/ies
the Jacobi
identity.

The bi-vector associated with $\hat{\mathcal{J}}_2$ is def\/ined by~\cite{olv2}
\begin{gather*}
\Theta_{\hat{\mathcal{J}}_2}=\frac12\int_{\mathbb R}(\vartheta\wedge\hat{\mathcal{J}}_2\vartheta){\rm d} x
=\int_{\mathbb R}\left(m\theta\wedge\theta_x+n\theta\wedge\zeta_x+n\zeta\wedge\theta_x-m\zeta\wedge\zeta_x\right),
\end{gather*}
where $\vartheta=(\theta,\zeta)$, $\theta$ and $\zeta$ denote the basic unit
vectors corresponding to $m$ and $n$, respectively, the notation $\wedge$
denotes the usual inner product between $\vartheta$ and $\hat{\mathcal{J}}_2\vartheta$.
It suf\/f\/ices to show that the Schouten bracket vanishes,
namely $[\hat{\mathcal{J}}_2,\hat{\mathcal{J}}_2]=0$.
In terms of
\begin{gather*}
\pr v_{\hat{\mathcal{J}}_2\vartheta}(m)=2m\theta_x+m_x\theta+2n\zeta_x+n_x\zeta,\qquad
\pr v_{\hat{\mathcal{J}}_2\vartheta}(n)=2n\theta_x+n_x\theta-2m\zeta_x-m_x\zeta,
\end{gather*}
a direct computation shows
\begin{gather*}
[\hat{\mathcal{J}}_1,\hat{\mathcal{J}}_2]=\pr v_{\hat{\mathcal{J}}_2v}(\Theta_{\hat{\mathcal{J}}_2})\\
\phantom{[\hat{\mathcal{J}}_1,\hat{\mathcal{J}}_2]}
=\int(2n\zeta_x\wedge\theta\wedge\theta_x+n_x\zeta\wedge\theta\wedge
\theta_x+2n\theta_x\wedge\theta\wedge\zeta_x-m_x\zeta\wedge\theta\wedge\zeta_x\\
\phantom{[\hat{\mathcal{J}}_1,\hat{\mathcal{J}}_2]=}
+n_x\theta\wedge\zeta\wedge\theta_x-2m\zeta_x\wedge\zeta\wedge\theta_x
-2m\theta_x\wedge\zeta\wedge\zeta_x-m_x\theta\wedge\zeta\wedge\theta_x){\rm d} x=0,
\end{gather*}
where the skew-symmetric property for the wedge product is used.

Next, we prove that the Hamiltonian operators $\hat{\mathcal{J}}_1$ and $\hat{\mathcal{J}}_2$ are
compatible, i.e.
\begin{gather*}
[\hat{\mathcal{J}}_1,\hat{\mathcal{J}}_2]+[\hat{\mathcal{J}}_2,\hat{\mathcal{J}}_1]
=\pr v_{\hat{\mathcal{J}}_1\vartheta}(\Theta_{\hat{\mathcal{J}}_2})
+\pr v_{\hat{\mathcal{J}}_2\vartheta}(\Theta_{\hat{\mathcal{J}}_1})=0.
\end{gather*}
Note that
\begin{gather*}
\pr v_{\hat{\mathcal{J}}_2\vartheta}(\Theta_{\hat{\mathcal{J}}_1})=0,\qquad
\pr v_{\hat{\mathcal{J}}_1\vartheta}(m)=\theta_{xxx}-\theta_x,\qquad
\pr v_{\hat{\mathcal{J}}_1\vartheta}(n)=\zeta_x-\zeta_{xxx}.
\end{gather*}
Through integration by parts, we get
\begin{gather*}
\big[\hat{\mathcal{J}}_1,\hat{\mathcal{J}}_2\big]+\big[\hat{\mathcal{J}}_2,\hat{\mathcal{J}}_1\big]
=\pr v_{\hat{\mathcal{J}}_1\vartheta}\big(\Theta_{\hat{\mathcal{J}}_2}\big)
+\pr v_{\hat{\mathcal{J}}_2\vartheta}\big(\Theta_{\hat{\mathcal{J}}_1}\big)\\
\hphantom{\big[\hat{\mathcal{J}}_1,\hat{\mathcal{J}}_2\big]+\big[\hat{\mathcal{J}}_2,\hat{\mathcal{J}}_1\big]}{}
=\int(\theta_{xxx}\wedge\theta\wedge\theta_x-\zeta_{xxx}\wedge\theta\wedge\zeta_x
+\zeta_{x}\wedge\zeta\wedge\theta_x-\zeta_{xxx}\wedge\zeta\wedge\theta_x\\
\hphantom{\big[\hat{\mathcal{J}}_1,\hat{\mathcal{J}}_2\big]+\big[\hat{\mathcal{J}}_2,\hat{\mathcal{J}}_1\big]=}{}
-\theta_{xxx}\wedge\zeta\wedge\zeta_x+\theta_{x}\wedge\zeta\wedge\zeta_x){\rm d} x\\
\hphantom{\big[\hat{\mathcal{J}}_1,\hat{\mathcal{J}}_2\big]+\big[\hat{\mathcal{J}}_2,\hat{\mathcal{J}}_1\big]}{}
=\int(-\zeta_{xxx}\wedge\theta\wedge\zeta_x-\zeta_{xxx}\wedge
\zeta\wedge\theta_x-\theta_{xxx}\wedge\zeta\wedge\zeta_x){\rm d} x\\
\hphantom{\big[\hat{\mathcal{J}}_1,\hat{\mathcal{J}}_2\big]+\big[\hat{\mathcal{J}}_2,\hat{\mathcal{J}}_1\big]}{}
=\int(-\zeta_{xxx}\wedge\theta\wedge\zeta_x-\zeta_{xxx}\wedge\zeta\wedge\theta_{x}
-\theta_{x}\wedge\zeta_x\wedge\zeta_{xx}-\theta_x\wedge\zeta\wedge\zeta_{xxx}){\rm d} x\\
\hphantom{\big[\hat{\mathcal{J}}_1,\hat{\mathcal{J}}_2\big]+\big[\hat{\mathcal{J}}_2,\hat{\mathcal{J}}_1\big]}{}
=\int(-\zeta_{xxx}\wedge\theta\wedge\zeta_x-\theta_x\wedge\zeta_x\wedge\zeta_{xx}){\rm d} x=0.
\end{gather*}
Thus $\hat{\mathcal{J}}_1$ and $\hat{\mathcal{J}}_2$ are a~Hamiltonian-pair.
Let's write the complex CH equation~\eqref{Two-CH} as follows
\begin{gather*}
\left(
\begin{matrix}m\\
n\end{matrix}\right)_t=\hat{\mathcal{J}}_2\left(
\begin{matrix}\dfrac{\delta\hat{H}_1}{\delta m}\vspace{1mm}\\
\dfrac{\delta\hat{H}_1}{\delta n}
\end{matrix}\right).
\end{gather*}
It is easy to get
\begin{gather*}
\frac{\delta\hat{H}_1}{\delta m}=-u,\qquad\frac{\delta\hat{H}_1}{\delta n}=v.
\end{gather*}
Thus
\begin{gather*}
\frac{\delta\hat{H}_1}{\delta u}=-m,\qquad\frac{\delta\hat{H}_1}{\delta v}=n.
\end{gather*}
Hence we deduce that
\begin{gather*}
\hat{H}_1=\frac12\int(vn-um){\rm d} x.
\end{gather*}
To compute $\hat{H}_2$, we write~\eqref{Two-CH} as
\begin{gather*}
\begin{split}
& m_t=-\partial\left(\frac32u^2-\frac12u_x^2-uu_{xx}-\frac32v^2+\frac12v_x^2+vv_{xx}\right),\\
& n_t=\partial\left(uv_{xx}+vu_{xx}+u_xv_x-3uv\right).
\end{split}
\end{gather*}
Since $\hat{H}_2$ satisf\/ies
\begin{gather*}
m_t=-\partial\big(1-\partial^2\big)\frac{\delta\hat{H}_2}{\delta m}=-\partial\frac{\delta\hat{H}_2}{\delta u},\qquad
n_t=\partial\big(1-\partial^2\big)\frac{\delta\hat{H}_2}{\delta n}=\partial\frac{\delta\hat{H}_2}{\delta v}.
\end{gather*}
It follows that
\begin{gather*}
\frac{\delta\hat{H}_2}{\delta u}=\frac32u^2-\frac12u_x^2-uu_{xx}-\frac32v^2+\frac12v_x^2+vv_{xx},\\
\frac{\delta\hat{H}_2}{\delta v}=uv_{xx}+vu_{xx}+u_xv_x-3uv.
\end{gather*}
Note that
\begin{gather*}
\frac32u^2-\frac12u_x^2-uu_{xx}=\frac{\delta}{\delta u}\frac12\int\left(u^3+uu_x^2\right){\rm d} x,\\
\frac{\delta}{\delta u}\int uv^2{\rm d} x=v^2,\\
\frac{\delta}{\delta v}\int uv^2{\rm d} x=2uv,\\
\frac{\delta}{\delta u}\int\left(-\frac12uv_x^2-u_xvv_x\right){\rm d} x=\frac12v_x^2+vv_{xx}.
\end{gather*}
Hence we arrive at
\begin{gather*}
\hat{H}_2=\frac12\int\left(u^3+uu_x^2-3uv^2-uv_x^2-2u_xvv_x\right){\rm d} x.\tag*{\qed}
\end{gather*}
\renewcommand{\qed}{}
\end{proof}

\section[Curve flows on $\mathbb{S}^n(1)$ and multi-component modif\/ied CH equations]{Curve f\/lows on $\boldsymbol{\mathbb{S}^n(1)}$ and multi-component\\ modif\/ied CH equations}\label{section5}

Assume that $\gamma(x,t)$ is a~curve f\/low on unit sphere
${\mathbb S}^n(1)={\rm SO}(n+1)/{\rm SO}(n)$, which satisf\/ies $\|\gamma\|=1$, where $x$ is
the invariant arc-length parameter, $t$ is the time.
The natural
frame of the curve $\gamma\in{\mathbb S}^n(1)$ is
$\{e_1=\gamma_x,e_2,\ldots,e_n\}$.
Let
$\rho=(e_0=\gamma,e_1,\ldots,e_n)\in {\rm SO}(n+1)$ be the lift from
${\mathbb S}^n(1)$ to bundle space ${\rm SO}(n+1)$, and $D_x$ and $D_t$
denote respectively the tangent and evolutionary vector f\/ield.
It
follows that
\begin{gather}\label{e5.1}
\rho_x=\rho\hat{\omega}(D_x),
\end{gather}
where $\hat{\omega}$ is the Cartan connection
\begin{gather*}
\hat{\omega}(D_x)=\left(
\begin{matrix}0&-1&\vec{0}^T\\
1&0&-\vec{k}^T\\
\vec{0}&\vec{k}&O
\end{matrix}
\right),
\qquad
O\in\mathfrak{so}(n-1),
\end{gather*}
(the natural frame formulae~\eqref{e5.1} for curves on the sphere
comes out from the Frenet formulae~\cite{dol} through the Hasimoto
transformation).
Here $\vec{k}=(k_1,k_2,\ldots,k_{n-1})$ is the
natural curvature vector of $\gamma$.

Assume that the curve f\/low is governed by
\begin{gather*}
\gamma_t=fe_1+h_1e_2+h_2e_3+\cdots+h_{n-1}e_n,
\end{gather*}
where the tangent velocity $f$ and normal velocities $h_i$ $(i=1,2,\ldots,n-1)$
depend on the curvatures and their derivatives with respect to arc-length $x$.

The induced time evolution for the frame is
\begin{gather*}
\rho_t=\rho\hat{\omega}(D_t),
\end{gather*}
with
\begin{gather*}
\hat{\omega}(D_t)=\left(
\begin{matrix}0&-f&-\vec{h}^T\\
f&0&-\vec{\xi}^T\\
\vec{h}&\vec{\xi}&\Theta
\end{matrix}
\right),
\qquad
\Theta\in\mathfrak{so}(n-1),
\end{gather*}
where $\vec{h},\vec{\xi}\in{\mathbb R}^{n-1}$, $\vec{\xi}$ is a
unknown vector, which will be determined later by the structure
equations.

First, we assume that the f\/low is intrinsic, namely, the
distribution $\{D_x,D_t\}$ satisf\/ies\linebreak
$[D_x,D_t]=0$ so that the
integral submanifold is a~smooth two-dimensional surface on Lie
group ${\rm SO}(n+1)$.
Making use of the Cartan structure equation
\begin{gather*}
\frac{d}{dt}\hat{\omega}(D_x)-\frac{d}{dx}\hat{\omega}(D_t)
-[\hat{\omega}(D_x),\hat{\omega}a(D_t)]=0,
\end{gather*}
one gets the following equations{\samepage
\begin{gather}
f_x=\langle \vec{k},\vec{h}\rangle ,\label{Constraint-5.1}\\
\vec{\xi}=\vec{h}_x+f\vec{k},\label{Constraint-5.2}\\
\Theta_x=\vec{k}\otimes\vec{\xi}-\vec{\xi}\otimes\vec{k},\label{Constraint-5.3}\\
\vec{k}_t=\vec{\xi}_x-\Theta
\vec{k}+\vec{h},\label{Constraint-5.4}
\end{gather}
where~\eqref{Constraint-5.1} is the arc-length preserving condition.}

For convenience, the following notations are used.
For any
$\vec{a},\vec{b}\in{\mathbb R}^{n-1}$, $\langle \vec{a},\vec{b}\rangle$
denotes the usual Euclidean inner product, i.e.,
$\langle \vec{a},\vec{b}\rangle =\vec{a}^T\vec{b}$, $\vec{a}\otimes\vec{b}$
denotes the tensor product, namely
\begin{gather*}
\vec{a}\otimes\vec{b}=\left(
\begin{matrix}a_1b_1&a_1b_2&\cdots&a_1b_{n-1}\\
a_2b_1&a_2b_2&\cdots&a_2b_{n-1}\\
\cdots&\cdots&\cdots&\cdots\\
a_{n-1}b_1&a_{n-1}b_2&\cdots&a_{n-1}b_{n-1}
\end{matrix}\right).
\end{gather*}
Def\/ine $\vec{a}\wedge\vec{b}=\vec{a}\otimes\vec{b}-\vec{b}\otimes\vec{a}$.
From~\eqref{Constraint-5.1} and \eqref{Constraint-5.2}, it follows that
\begin{gather}\label{e5.6}
f=\partial_x^{-1}\langle \vec{k},\vec{h}\rangle,\qquad
\vec{\xi}=\vec{h}_x+(\partial_x^{-1}\langle \vec{k},\vec{h}\rangle)\vec{k}.
\end{gather}
In view of~\eqref{Constraint-5.3}, we have
\begin{gather}\label{e5.7}
\Theta=\partial_x^{-1}(\vec{k}\wedge\vec{\xi}).
\end{gather}
Plugging~\eqref{e5.7} and \eqref{e5.6} into \eqref{Constraint-5.4} leads to the equation for the
curvature vector
\begin{gather}\label{e5.8}
\vec{k}_t=\vec{h}_{xx}+\langle \vec{k},\vec{k}\rangle \vec{h}+\big(\partial_x^{-1}\langle \vec{k},
\vec{h}\rangle \big)\vec{k}_x+\big(\partial_x^{-1}(\vec{k}_x\wedge\vec{h})\big)\vec{k}+\vec{h},
\end{gather}
where the identity for vectors
\begin{gather*}
(\vec{a}\wedge\vec{b})\cdot\vec{c}=\langle \vec{b},\vec{c}\rangle
\vec{a}-\langle \vec{a},\vec{c}\rangle \vec{b}
\end{gather*}
was used.

Analogous to the derivation for the CH equation~\cite{fuc2} and the modif\/ied
CH equation~\cite{olv3}, we restrict our attention to the following cases.

  {\bf Case 1.} $\vec{h}=\vec{u}_x$, $\vec{k}=\tilde{m}=\vec{u}+\vec{u}_{xx}$.
In this case, the tangent velocity $f$ is determined by
\begin{gather}\label{e5.9}
f=\partial_x^{-1}\langle \vec{u}+\vec{u}_{xx},\vec{u}_x\rangle
=\frac12\big(|\vec{u}|^2+|\vec{u}_x|^2\big)+c_0
\equiv\frac12\tilde{Q}+c_0,
\end{gather}
where $c_0$ is an integration constant.
Substituting~\eqref{e5.9} with $c_0=-1$
together with the expressions for $\vec{k}$ and $\vec{h}$ into~\eqref{e5.8}, and noting that
\begin{gather*}
\big(\partial_x^{-1}(\vec{k}_x\wedge\vec{h})\big)\vec{k}
=\left(\partial_x^{-1}(\vec{u}_{xxx}\wedge\vec{u}_x)\right)\tilde{m}
=\left(\vec{u}_{xx}\wedge\vec{u}_x\right)\tilde{m}\\
\phantom{\big(\partial_x^{-1}(\vec{k}_x\wedge\vec{h})\big)\vec{k}}
=-\left(\vec{u}\wedge\vec{u}_x\right)\tilde{m}-\left(\vec{u}_x\wedge\tilde{m}\right)\tilde{m}
=-\left(\vec{u}\wedge
\vec{u}_x\right)\tilde{m}-\langle  \tilde{m},\tilde{m}\rangle \vec{u}_x+\frac12\tilde{Q}_x\tilde{m},
\end{gather*}
we arrive at the multi-component modif\/ied CH equation
\begin{gather}\label{H-M-CH}
\tilde{m}_t=\frac12\big(\tilde{m}\tilde{Q}\big)_x-(\vec{u}\wedge\vec{u}_x)\tilde{m}.
\end{gather}

Thus we have established the following result.

\begin{theorem}
Assume that curves $\gamma(x,t)$ on the sphere ${\mathbb S}^n(1)$ $(n\geq1)$ are governed by the
flow
\begin{gather}\label{HM-CH-Flow}
\gamma_t=\left(\frac12\tilde{Q}-1\right)e_1+\sum\limits_{j=1}^{n-1}u_{j,x}e_{j+1},
\end{gather}
where $\{e_1,e_2,\ldots,e_n\}$ is the natural frame of the curve $\gamma(x,t)$, $(u_1,u_2,\ldots,
u_{n-1})$
is defined by the
curvatures $\vec{k}=\tilde{m}=\vec{u}+\vec{u}_{xx}$, $\tilde{Q}=|\vec{u}|^2+|\vec{u}_x|^2$.
Then the flow~\eqref{HM-CH-Flow} is intrinsic and the curvature vector $\tilde{m}$ fulf\/ills the
equation~\eqref{H-M-CH}.
\end{theorem}

\begin{remark}
{\rm In the case of $n=2$, i.e., the case of ${\mathbb S}^2(1)$, equation~\eqref{H-M-CH} reduces to the scaler modif\/ied CH equation
\eqref{e1.3} with $\delta=-1$, which is completely integrable.
In the
case of $n=3$, let $u_1=u$, $u_2=v$, $m=u+u_{xx}$, $n=v+v_{xx}$, then
system~\eqref{H-M-CH} reduces to
\begin{gather*}
m_t=\frac12\left[\big(u^2+v^2+u_x^2+v_x^2\big)m\right]_x-(uv_x-vu_x)n,\nonumber\\
n_t=\frac12\left[\big(u^2+v^2+u_x^2+v_x^2\big)n\right]_x+(uv_x-vu_x)m.
\end{gather*}
In general, the multi-component system~\eqref{H-M-CH} can be written as
\begin{gather*}
m_{i,t}=\frac12\sum\limits_{j=1}^{n-1}
\left[\big(u_j^2+u_{j,x}^2\big)m_i\right]_x\!-\!\sum\limits_{j=1}^{n-1}(u_iu_{j,x}\!-\!u_ju_{i,x})m_j,
\quad
1\leq i\leq n\!-\!1,
\quad
m_i=u_i+u_{i,xx}.
\end{gather*}}
\end{remark}

 {\bf Case 2.} $\vec{h}=\vec{u}_x$, $\vec{k}=\vec{m}=\vec{u}-\vec{u}_{xx}$.
In this case, the tangent velocity $f$ is given by
\begin{gather}\label{a5.9}
f=\partial_x^{-1}\langle \vec{u}-\vec{u}_{xx},\vec{u}_x \rangle
=\frac12\sum\limits_{i=1}^{n-1}\big(u_i^2-u_{i,x}^2\big)+c_1
\equiv\frac12Q+c_1,
\end{gather}
where $c_1$ is an integration constant.
Substituting~\eqref{a5.9} with $c_1=1$ into \eqref{e5.8} and noting that
\begin{gather*}
\big(\partial_x^{-1}(\vec{k}_x\wedge\vec{h})\big)\vec{k}
=-\left(\vec{u}\wedge\vec{u}_x\right)\vec{m}-\left(\vec{u}_x\wedge\vec{m}\right)\vec{m}
=-\left(\vec{u}\wedge\vec{u}_x\right)\vec{m}-\langle \vec{m},\vec{m}\rangle \vec{u}_x+\frac12Q_x\vec{m},
\end{gather*}
we obtain the multi-component modif\/ied CH equation
\begin{gather}\label{M-CH}
\vec{m}_t=\frac12\left(\vec{m}Q\right)_x-(\vec{u}\wedge\vec{u}_x)\vec{m}
+2\vec{u}_x,\qquad\vec{m}=\vec{u}-\vec{u}_{xx}.
\end{gather}

Thus we have proved the following result.

\begin{theorem}
Assume that curves $\gamma(x,t)$ on the sphere ${\mathbb S}^n(1)$ $(n\geq1)$ are governed by the
flow
\begin{gather}\label{M-CH-Flow}
\gamma_t=\left(\frac12Q+1\right)e_1+\sum\limits_{j=1}^{n-1}u_{j,x}e_{j+1},
\end{gather}
where $\{e_1,e_2,\ldots,e_n\}$ is the natural frame of the curves
$\gamma(x,t)$, $(u_1,u_2,\ldots,u_{n-1})$
is defined by the curvatures $\vec{k}=\vec{m}=\vec{u}-\vec{u}_{xx}$,
$Q=|\vec{u}|^2-|\vec{u}_x|^2$.
Then the flow~\eqref{M-CH-Flow}
is intrinsic and the curvature vector $\vec{m}$ satisfies~\eqref{M-CH}.
\end{theorem}

  {\bf Case 3.} $\vec{h}=\vec{u}_x$, $\vec{k}=\vec{m}=-\vec{u}_{xx}$.
In this case, the tangent velocity $f$ is determined by
\begin{gather*}
f=\partial_x^{-1}\langle -\vec{u}_{xx},\vec{u}_x\rangle =-\frac12|\vec{u}_x|^2+c_2,
\end{gather*}
where $c_2$ is an integration constant.
It is inferred from
\eqref{Constraint-5.2} and~\eqref{Constraint-5.3} that
\begin{gather*}
\vec{\xi}=\left(\frac12|\vec{u}_x|^2+1-c_2\right)\vec{u}_{xx},\qquad
\Theta_x=0.
\end{gather*}
Setting $\Theta=0$, $c_2=1$ and $\vec{v}=\vec{u}_x$, we arrive at the multi-component short-pulse
equation
\begin{gather}\label{M-SP}
\vec{v}_{xt}+\frac12\big(|\vec{v}|^2\vec{v}_x\big)_x+\vec{v}=0.
\end{gather}

Hence we could prove the following result.

\begin{theorem}
Assume that curves $\gamma(x,t)$ on the sphere ${\mathbb S}^n(1)$ $(n\geq1)$ are governed by the
flow
\begin{gather}\label{SP-Flow}
\gamma_t=\left(1-\frac12|\vec{u}_x|^2\right)e_1+\sum\limits_{j=1}^{n-1}u_{j,x}e_{j+1},
\end{gather}
where $\{e_1,e_2,\ldots,e_n\}$ is the natural frame of the curves
$\gamma(x,t)$, $(u_1,u_2,\ldots,u_{n-1})$
is defined by the curvatures $\vec{k}=\vec{m}=-\vec{u}_{xx}$.
Then the flow~\eqref{SP-Flow} is intrinsic and the curvature vector $\vec{v}=\vec{u}_x$
satisfies~\eqref{M-SP}.
\end{theorem}

The scalar equation of~\eqref{M-SP} was derived by Sch\"{a}fer
and Wayne~\cite{sw} as a~model for the propagation of ultra-short light
pulses in silica optical f\/ibers, which is also an approximation of nonlinear
wave packets in dispersive media in the limit of few cycles on the ultra-short pulse scale.

It is worth noting that, analogous to the system~\eqref{H-M-CH},
there is another version of multi-component generalization of
the modif\/ied Camassa--Holm equation~\eqref{e1.3} with $\delta=1$
\begin{gather}\label{M-M-CH}
\vec{m}_t=\frac12(\vec{m}Q)_x-(\vec{u}\wedge
\vec{u}_x)\vec{m},\qquad\vec{m}=\vec{u}-\vec{u}_{xx},
\end{gather}
(where $Q=|\vec{u}|^2-|\vec{u}_x|^2$).
However, in contrast to the
system~\eqref{H-M-CH} arising from the compact Riemannian symmetric
space ${\mathbb S}^n(1)={\rm SO}(n+1)/{\rm SO}(n)$ with positive constant
curvature, the system~\eqref{M-M-CH} arises from the noncompact
Riemannian symmetric space with negative constant curvature, i.e.,
the hyperbolic space ${\mathbb H}^n={\rm SO}(n,1)/{\rm SO}(n)$.
The derivation to
integrable curve f\/lows is similar to that for the ${\mathbb S}^n(1)$ case.
In fact, the Lie algebra structure corresponding to the hyperbolic space ${\mathbb
H}^n={\rm SO}(n,1)/{\rm SO}(n)$ is described by
\begin{gather*}
\mathfrak{so}(n,1)={\mathfrak h}\oplus{\mathfrak
m}=\mathfrak{so}(n)\oplus{\mathbb R}^n,
\end{gather*}
with
\begin{gather*}
\left(
\begin{matrix}
0&p^T\\
p&0\\
\end{matrix}
\right)\in{\mathfrak m},\qquad\left(
\begin{matrix}
0&0\\
0&\Theta\\
\end{matrix}
\right)\in{\mathfrak h}
\end{gather*}
where $p\in{\mathbb R}^n$, $\Theta\in\mathfrak{so}(n)$.
The
Cartan connection matrices $\hat{\omega}(D_x)$ and $\hat{\omega}(D_t)$ of the natural frame are
replaced with
\begin{gather*}
\hat{\omega}(D_x)=\left(
\begin{matrix}
0&1&\vec{0}^T\\
1&0&-\vec{k}^T\\
\vec{0}&\vec{k}&O
\end{matrix}
\right),
\qquad
O\in\mathfrak{so}(n-1)
\end{gather*}
and
\begin{gather*}
\hat{\omega}(D_t)=\left(
\begin{matrix}
0&f&\vec{h}^T\\
f&0&-\vec{\xi}^T\\
\vec{h}&\vec{\xi}&\Psi
\end{matrix}
\right),
\qquad
\Psi\in\mathfrak{so}(n-1),
\end{gather*}
for curves on ${\mathbb H}^n={\rm SO}(n,1)/{\rm SO}(n)$.
Similar results can be derived for
the Hyperbolic space~${\mathbb H}^n$.
Integrability of the equations~\eqref{H-M-CH} and~\eqref{M-M-CH} are guaranteed by
the following results.

\begin{theorem}
The systems~\eqref{M-M-CH} and \eqref{H-M-CH} are Lax integrable, namely
they admit the following $(n+1)\times(n+1)$ Lax-pair
\begin{gather*}
\phi_x=U\phi,\qquad\phi_t=V\phi,
\end{gather*}
where for~\eqref{M-M-CH},
\begin{gather*}
U=\left(
\begin{matrix}
0&1&\vec{0}^T\\
1&0&\lambda \vec{m}^T\\
\vec{0}&-\lambda \vec{m}&0
\end{matrix}
\right),\qquad
V=\left(
\begin{matrix}
0&\frac 12 Q+\lambda^{-2}&\lambda^{-1}\vec{u}_x^T\\
\frac 12 Q+\lambda^{-2}&0&\lambda^{-1} \vec{u}^T+\frac 12 \lambda Q\vec{m}^T\\
\lambda^{-1}\vec{u}_x&-\lambda^{-1}\vec{u}-\frac 12 \lambda Q \vec{m}&-\vec{u}\wedge \vec{u}_x
\end{matrix}
\right),
\end{gather*}
and for~\eqref{H-M-CH},
\begin{gather*}
U=\left(\!
\begin{matrix}
0&1&\vec{0}^T\\
-1&0&\lambda \tilde{m}^T\\
\vec{0}&-\lambda \tilde{m}&0
\end{matrix}\!
\right),\qquad
V=\left(\!
\begin{matrix}
0&\frac 12 \tilde{Q}-\lambda^{-2}&\lambda^{-1}\vec{u}_x^T\\
-\frac 12 \tilde{Q}+\lambda^{-2}&0&-\lambda^{-1} \vec{u}^T+\frac 12 \lambda \tilde{Q}\tilde{m}^T\\
-\lambda^{-1}\vec{u}_x&\lambda^{-1}\vec{u}-\frac 12 \lambda \tilde{Q} \tilde{m}&-\vec{u}\wedge \vec{u}_x
\end{matrix}\!
\right).
\end{gather*}
\end{theorem}

\section{Concluding remarks}\label{section6}

In this paper, geometrical formulations to several multi-component
integrable systems are provided.
These systems are regarded as
multi-component generalizations of the CH equation and the modif\/ied CH
equation, which can be obtained through the tri-Hamiltonian duality approach.
We showed that an integrable generalization to the nonlinear
Schr\"{o}dinger equation arises from a~non-stretching invariant
curve f\/low in the three-dimensional Euclidean geometry.
The
integrable complex CH equation comes from an invariant curve f\/low on
the M\"{o}bius 2-sphere.
Furthermore, we verif\/ied that
multi-component generalizations to the modif\/ied CH equation arise
naturally from the curve f\/lows in $n$-dimensional sphere ${\mathbb
S}^n(1)$ and the hyperbolic space ${\mathbb H}^n={\rm SO}(n,1)/{\rm SO}(n)$.

In~\cite{kam}, Olver, Kamran and Tenenblat have established the theory for curves in af\/f\/ine
symplectic
geometry.
The curve f\/lows in four-dimensional af\/f\/ine symplectic geometry were studied in~\cite{val},
and an integrable three-component equation with bi-Hamiltonian structure was obtained.
The theory for curves
in the centro-equiaf\/f\/ine symplectic geometry was established in~\cite{son1}.
It was shown that certain invariant
curve f\/lows in the centro-equiaf\/f\/ine symplectic geometry yield noncommutative KdV
equations~\cite{olv4}.
It is
still not clear that what are the dual version of these integrable equations arising from curve
f\/lows in the af\/f\/ine
and centro-equiaf\/f\/ine symplectic geo\-metries.

\subsection*{Acknowledgements} The authors would like to thank the anonymous
referees for constructive suggestions and comments.
This work was supported by the China NSF for Distinguished Young Scholars
under Grant 10925104 and the China NSF under Grants 11071278 and 60970054.

\pdfbookmark[1]{References}{ref}
\LastPageEnding

\end{document}